\documentclass[twocolumn,floats,showpacs,amsmath,amssymb,prb]{revtex4}
\usepackage{graphicx}
\begin{document}

\title{Path-integral molecular dynamics simulation of 3C-SiC}
\author{Rafael Ram\'{\i}rez}
\author{Carlos P. Herrero}
\affiliation{Instituto de Ciencia de Materiales de Madrid,
         Consejo Superior de Investigaciones Cient\'{\i}ficas (CSIC),
         Campus de Cantoblanco, 28049 Madrid, Spain }
\author{Eduardo R. Hern\'andez}
\affiliation{Institut de Ci\`encia de Materials de Barcelona (ICMAB),
         Consejo Superior de Investigaciones Cient\'{\i}ficas (CSIC),
         Campus de Bellaterra, 08193 Barcelona, Spain }
\author{Manuel Cardona}
\affiliation{Max-Planck-Institut f\"ur Festk\"orperforschung,
         Heisenbergstrasse 1, 70569 Stuttgart, Germany}
\date{\today}

\begin{abstract}
Molecular dynamics simulations of 3C-SiC have been performed
as a function of pressure and temperature.
These simulations treat both electrons and atomic nuclei by quantum
mechanical methods. 
While the electronic structure of the solid 
is described by an efficient tight-binding 
Hamiltonian, the nuclei dynamics is treated by the path
integral formulation of statistical mechanics.
To assess the relevance of nuclear quantum effects,  
the results of quantum simulations are compared to others
where either the Si nuclei, the C nuclei or both atomic nuclei
are treated as classical particles.
We find that the experimental thermal expansion of 3C-SiC is 
realistically reproduced by our simulations.
The calculated bulk modulus of 3C-SiC and its pressure derivative at room
temperature show also good agreement with the available experimental data.
The effect of the electron-phonon interaction 
on the direct electronic gap of 3C-SiC has been calculated
as a function of temperature
and related to results obtained for bulk diamond and Si.
Comparison to available experimental data 
shows satisfactory agreement, although we observe that the employed 
tight-binding model tends to overestimate the magnitude of 
the electron-phonon interaction.
The effect of treating the atomic nuclei as classical particles 
on the direct gap of 3C-SiC has been assessed. 
We find that non-linear quantum effects related 
to the atomic masses are particularly
relevant at temperatures below 250 K.
\end{abstract}

\pacs{63.20.Kr,05.30.-d,71.15.Pd}

% 63.20.Kr Phonon-electron and phonon-phonon interactions 
% 05.30.-d Quantum statistical mechanics
% 71.15.Pd Molecular dynamics calculations (Car-Parrinello) and other 
%  numerical simulations

\maketitle

\section{Introduction}
Silicon carbide has attracted much interest in technology
as a wide-gap semiconductor in high power electronic devices.
\cite{ma06}
It features large electronic band gaps, extreme hardness, 
large thermal conductivity and excellent chemical stability.
It is a prototype of materials which exhibit several polytypes
(more than 200) and the only IV-IV compound which possesses long range order
polytypes.
The cubic modification 3C-SiC has a zinc-blende structure, 
being thus the type with closest structural
relationship to both diamond and elemental Si.
From the point of view of basic research, 
there exists a large body of experimental data on silicon carbide, 
so that the quality of theoretical approaches and computer simulations
can be contrasted against them.\cite{yo02}

The influence of anharmonic effects in the
vibrational properties of 3C-SiC is revealed
by several experimental studies.
The pressure dependence of optical phonons in 3C-SiC    
has been measured by means of first- and 
second-order Raman scattering up to 23 GPa,
\cite{ol82a}
while their temperature dependence 
has been studied by first-order Raman scattering up to 750 K.
\cite{ol82b}
The pressure dependence of phonon lifetimes,
which provides detailed information on the anharmonic phonon-phonon 
interaction, has been also
calculated from first principles using 
perturbation theory.\cite{de01,ka96}
These calculations permitted a detailed analysis of the
microscopic anharmonic mechanism and yielded accurate predictions
of experimental properties.

The thermal expansion of the lattice is another property
determined by the anharmonicity of the interatomic potential.
There is no evidence of negative thermal expansion in 3C-SiC,
in contrast to diamond and elemental Si, which show
negative thermal expansion at temperatures below 100 K. \cite{bi89,sa02}
An analysis of the most reliable experimental data for the 
thermal expansion of 3C-SiC led to a tabulation of 
recommended values for this quantity in Ref. \onlinecite{sl75}.
Theoretical results for the temperature dependence of the linear thermal
expansion for 3C-SiC have been also reported using the
quasiharmonic approximation.\cite{tal95,ka94}

The optical properties of 3C-SiC have been measured using spectroscopic
ellipsometry as a function of temperature in the range
between 90 and 550 K.\cite{pe98}
The electron-phonon interaction
is responsible for the decrease found in the interband transition
energies as temperature increases.
We recall that the renormalization of the optical response of
semiconductors by electron-phonon interaction
has been a topic of increasing interest in recent 
years,\cite{ca01,ca05} with special focus in the characterization
of isotopic effects.\cite{ca05r}
From a theoretical point of view,
the electron-phonon interaction in tetrahedral semiconductors
has been studied so far by perturbation theory.
\cite{al81,zo92,ki89}      

An interesting alternative to perturbational approaches to study
the coupling between electronic and vibrational 
degrees of freedom in solids, is the 
combination of the path integral (PI) formulation
with electronic structure methods.
The path integral approach to statistical mechanics
allows us to study finite temperature
properties that are related to the quantum nature of the
atomic nuclei. \cite{fe72,ce95}
An advantage of its combination with electronic structure methods
is that both the electrons and the atomic
nuclei are then treated quantum mechanically in the framework of the
Born-Oppenheimer (BO) approximation, so that 
phonon-phonon and electron-phonon interactions
are automatically considered in the simulation.    
This unified scheme has been applied so far to the study of
solids and molecules containing light atoms.
\cite{marx96,tu97,ra98,si01,mo01,ch03,sa04,oh04}  
A recent application of this method has shown that
the electron-phonon coupling leads to a zero-point renormalization of
the direct electronic gap of diamond of 10\%,\cite{ra06}
in agreement with a previous perturbational analysis.\cite{zo92}   

In this paper we present a path integral molecular dynamics
study of 3C-SiC at temperatures between 100 and 1200 K and
pressures up to 60 GPa.
The electronic structure was treated with a non-orthogonal
tight-binding (TB) Hamiltonian as a reasonable compromise to
reduce the computational cost of deriving the BO energy surface for the
nuclear dynamics.
We are interested in the simulation of vibrational    
properties that rely on phonon-phonon
interactions, such as the temperature dependence
of the linear expansion coefficient, and also in the simulation
of electronic properties that are determined by electron-phonon
coupling, such as the temperature dependence of the direct electronic gap.

This paper is organized as follows. In Sec.\,II, we describe the
computational method employed in our simulations.
Our results are presented and discussed in Sec.\,III, dealing with the
thermal expansion coefficient and the temperature dependence of the
direct electronic gap of 3C-SiC. 
The results for the electronic gap in 3C-SiC will be then
related to those derived for diamond and crystalline Si.  
The results of the simulation will also be compared to available
experimental data.
The pressure dependence of the electronic gap as well as
the results for the bulk modulus and its pressure derivative
at 300 K will complete Sec.\,III.     
In Sec.\,IV, we present the main conclusions of the paper.     

\section{Computational Method}

The formalism employed here for the quantum treatment of 
electrons and nuclei is based on the combination of the
path integral formulation, to derive properties
of the atomic nuclei in thermal equilibrium,
with an electronic tight-binding Hamiltonian to describe the
BO energy surface, $E_{BO}({\bf R})$, of 3C-SiC
as a function of the nuclear configuration ${\bf R}$. 
This approach has been recently used in our
simulation of diamond\cite{ra06}
and isolated hydrogenic impurities in diamond\cite{he06} in the
canonical $NVT$ ensemble (number of atoms $N$, volume $V$, and
temperature $T$ are constant).
Therefore, we present here only a brief summary of the method,
with focus on those extensions required for the 3C-SiC
simulations, that were performed in both the $NVT$ and
the isothermal-isobaric $NPT$ ensemble ($N$, pressure $P$, and $T$ 
are constant).
The combination of the path integral formalism with ab initio 
Hamiltonians based on density funtional theory (DFT) is an interesting
alternative that has been reviewed in the literature, \cite{ma00,tu02b}
but it has not been applied so far to the investigation of the
temperature dependence of the optical response in semiconductors.
Typically tight binding methods are two orders of magnitude faster
than ab initio DFT methods.

The employed tight-binding one-electron effective Hamiltonian is
based on density functional (DF) calculations.\cite{po95}
The TB energy consists of two terms, the first one of which is the 
sum of occupied one-electron state energies, and the second is given
by a pair-wise repulsive interatomic potential.
The one-electron states are derived by diagonalizing a
two-center Hamiltonian using a minimal basis of
non-orthogonal atomic orbitals. The pair potential is adjusted so that
the DF energy is reproduced for a series of reference systems (e.g. the
dimer, the crystal, etc.) The total energy is thus obtained
by adding to the electronic energy the repulsive pair-potential.
Preliminary calculations on 3C-SiC revealed that the original
parameterization of the pair-potential between Si and C leads
to an overestimation of anharmonic effects in 3C-SiC,
a fact that motivated us to improve the parameterization
of the Si-C pair-potential as explained in Appendix \ref{app1}.       

The computational advantage of using the path-integral formulation of 
statistical mechanics is based on
the so-called ``quantum-classical'' isomorphism.
Thus, this method exploits the fact that
the partition function of a quantum system is formally equivalent to
that of a classical one, obtained by replacing each quantum particle 
(here, atomic nucleus) by a
ring polymer consisting of $L$ ``beads'', connected by harmonic
springs.\cite{gi88,ce95,fe72,kl90}
In many-body problems, the configuration space of the classical
isomorph is usually sampled by
Monte Carlo or molecular dynamics (MD) techniques. 
Here, we have employed the PI MD method, 
which has been found to require less computer time resources
when applied to our problem.
Effective algorithms to perform PI MD simulations in the canonical $NVT$
ensemble have been described in detail by Martyna {\em et al.}\cite{ma96} 
and by Tuckerman.\cite{tu02}
The extensions for PI MD simulations in the $NPT$ ensemble
require the definition of appropriate dynamical equations
of motion to include the volume as fluctuating dynamical
variable, and multiple time step algorithms to integrate
these equations numerically. 
These extensions are well documented in Refs. \onlinecite{ma99,tu98,tu06}.
For the simulation of 3C-SiC in the $NPT$ ensemble
only isotropic volume fluctuations were allowed.
All calculations presented here were carried out 
using originally developed software,
which enables efficient PI MD simulations on parallel supercomputers.

Simulations were performed on a $2\times2\times2$ supercell of the
3C-SiC face-centered cubic cell with periodic boundary conditions,
containing $N=$ 64 atoms. 
Convergence criteria used previously for diamond were
proved to apply also for the 3C-SiC simulations.
Thus, we used only the $\Gamma$ point for the sampling of the Brillouin 
zone~(BZ) of the simulation supercell in the electronic structure calculation. 
A set of 4 ${\bf k}$ points would increase the computer time
by a factor of 10 without significant changes
of the results presented here.      
For a given temperature, a typical run consisted of $10^4$ MD steps for
system equilibration, followed by $10^5$ steps for the calculation of 
ensemble average properties.
To have a nearly constant precision in the path integral results
at different temperatures, we have taken a number of 
beads, $L$ (Trotter number), that scales with 
$LT = 6000$ K. 
The atomic masses of C and Si were set to 12.011 and 28.086 amu,
respectively.
Within the employed formalism,
the classical limit of a given nucleus is reached 
by setting the corresponding nuclear mass tending to infinity.
Thus, for comparison with the results of our full PI MD simulations, 
we have carried out some PI MD simulations were the mass of either
C or Si was set to a very large number ($8 \times 10^4$ amu).
The main effect of setting such a large nuclear mass in the
path integral simulation is that the ring polymer
associated to the atomic nucleus shrinks and looks just like a point
(classical) particle. Moreover, the calculation of thermostatted equations
of motion by using chains of Nos\'e-Hoover thermostats (see below) ensures
that the canonical probability distribution of the shrinked classical-like
particles is correctly sampled as a function of temperature.  
Also classical MD simulations were performed with
the same interatomic interaction.
The classical limit is easily achieved within the PI algorithm
by setting the Trotter number $L$ = 1.

The quantum simulations were performed using a staging transformation
for the bead coordinates.
Chains of four Nos\'e-Hoover thermostats 
were coupled to each of the staging variables to generate 
the canonical $NVT$ ensemble.
\cite{tu98}
To integrate the equations of motion we have used
the reversible reference system propagator algorithm (RESPA), which allows
one to define different time steps for the integration of the fast and slow
degrees of freedom.\cite{ma96}
For the evolution of the fast dynamical variables, that include the
thermostats and harmonic bead interactions, we used a
time step $\delta t = \Delta t/4$, where
$\Delta t$ is the time step associated to the calculation of TB forces.
A value of $\Delta t$ = 0.5~fs was found to provide 
adequate convergence.
The thermostat ``mass'' parameter, $Q$, is chosen to evolve in the
scale of the harmonic bead forces by being defined as\cite{tu98}
\begin{equation}
Q = \frac{\beta \hbar^2}{L}  \,
\label{Q}
\end{equation}
where $\beta$ = $(k_B T)^{-1}$ represents the inverse temperature.
In the case of the $NPT$ ensemble a chain of four barostats
was coupled to the volume and the barostat ``mass'' parameter
was set 100 times larger than the thermostat one. 
By calculating the average pressure in the $NVT$ ensemble we could 
check the internal consistency between our $NPT$ and $NVT$ simulations.
The pressure estimator used in the $NVT$ ensemble is 
given in Appendix \ref{app2}.
The direct electronic gap of 3C-SiC was derived as 
\begin{equation}
E_0 = \langle E_c \rangle - \langle E_v \rangle  \, ,
\end{equation}  
where $\langle E_c \rangle$ and $\langle E_v \rangle$ are the 
ensemble average of the one-electron states associated with the 
bottom of the conduction band and the top of the valence band 
at the $\Gamma$ (${\bf k = 0}$) reciprocal lattice point. 
For details on the calculation of the expectation values 
$\langle E_c \rangle$ and $\langle E_v \rangle$ see
Ref. \onlinecite{ra06}.

Several sets of simulations were performed in this work.
Two sets of 3C-SiC simulations
were done at temperatures between 100 and 1200 K.
One set corresponds to $NPT$ simulations at constant pressure ($P$=0),
and the other to $NVT$ simulations at constant volume ($a=4.3594$ \AA).
Another set of $NPT$ simulations was made at room temperature ($T=300$ K)
and pressures in the range between -10 and 60 GPa.    
Additional constant volume simulations were done for
diamond and Si as a function of temperature.

\begin{figure}
\vspace{-3.8cm}
\includegraphics[width= 9cm]{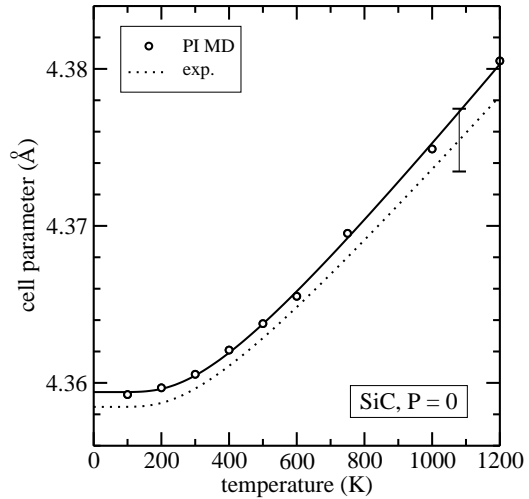}
\vspace{-1.5cm}
\caption{
Temperature dependence of the cell parameter of 3C-SiC.
Open circles are the results of PI MD simulations
at zero pressure.
The continuous line is a fit to Eq. (\ref{fit}).
The dotted line is the result derived from the experimental values of the
thermal expansion coefficients given in Ref. \onlinecite{sl75}.
The error bar is derived from the uncertainty of the experimental data.
The statistical error of the simulation results is of the size of the
symbols.
}
\label{fig1}
\end{figure}

\section{Results and Discussion}

The results derived from our PI MD simulations are presented
in the next subsections. Firstly, we focus on the temperature dependence
of the cell parameter and the linear expansion coefficient at zero pressure.

\subsection{Thermal expansion}

\begin{figure}
\vspace{-3.8cm}
\includegraphics[width= 9cm]{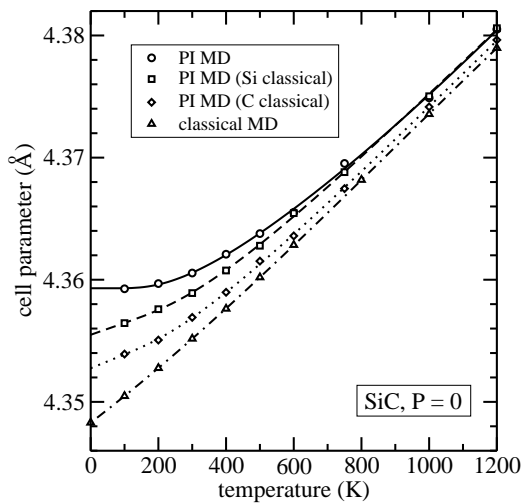}
\vspace{-1.5cm}
\caption{
The temperature dependence of the cell parameter of 3C-SiC
(circles) derived from PI MD simulations is compared to PI MD simulations
performed by selectively setting either the Si nuclei (squares)
or the C nuclei (diamonds) as classical particles.
The results of classical MD simulations are shown by triangles.
The lines represent empirical fits as explained in the text.
The statistical error of the simulation results is of the size of the
symbols.
}
\label{fig2}
\end{figure}

The temperature dependence of the cell parameter, $a(T)$, of 3C-SiC 
at zero pressure is presented in Fig. \ref{fig1}.
The simulation results are plotted as open circles.
The dotted line is derived from the
recommended experimental values of the linear expansion
coefficient of 3C-SiC by a numerical integration.\cite{sl75}
The resulting relative changes in the cell parameter, $\Delta a(T)$, 
were converted to an absolute
scale by using the experimental value of $a=4.3596$ \AA\ at 297 K.
\cite{la82}
Both simulation and experimental data were fitted with
a standard Bose-Einstein expression\cite{ca05}
\begin{equation}
a(T) = a_0 + b \left(1 + \frac{2}{\exp(\Theta_a/T) -1} \right)  \,\,  ,
\label{fit}
\end{equation}
the values of the fitted parameters are summarized in Table \ref{tab1}.
The results in Fig. \ref{fig1} show that
the simulations and experimental data agree
within the uncertainty of the experimental values.
At temperatures below 600 K the calculated cell parameter $a(T)$ shows 
a rigid shift with respect to experiment
of about $10^{-3}$ \AA\ toward larger values.
This shift increases slightly at higher temperatures.
The most plausible explanation for this high temperature deviation
of simulated data from experiment is that the employed TB model
overestimates the anharmonicity of the interatomic potential
when the amplitude of nuclei vibrations is activated thermally to 
values significantly larger than low temperature zero-point motions.  

% -----------------------------------------------------------------
\begin{table}
\caption{
Values of the parameters obtained by fitting the temperature
dependence of the lattice constant, $a(T)$, of 3C-SiC with Eq. (\ref{fit}).
The results correspond to $NPT$ simulations at $P=0$.
The experimental parameters were derived from the recommended
thermal expansion coefficients of Ref. \onlinecite{sl75}.
The PI MD parameters for those simulations where the Si or C nuclei are
treated in the classical limit were obtained by adding a linear term,
$a_1T$,
to Eq. (\ref{fit}).
The classical MD parameters correspond to the cubic function given in
Eq. (\ref{fit_cla}).
}
~\newline
\begin{tabular}{|l|c|c|c|c|}
\hline
               &$a_0$ (\AA)& $a_1$ (\AA K$^{-1})$ & $b$(\AA)
&$\Theta_a$(K)\\
\colrule
PI MD          &  4.3463   &   -          & 0.0131       &     975.7 \\
exp.           &  4.3475   &   -          & 0.0110       &     898.0 \\
PI MD (Si cla.)&  4.3458   & 1.0 10$^{-5}$& 0.0097       &    1077.2 \\
PI MD (C  cla.)&  4.3451   & 1.1 10$^{-5}$& 0.0077       &     921.7 \\
\hline
                &$a_0$ (\AA)& $a_1$ (\AA K$^{-1}$)& $a_2$ (\AA
  K$^{-2})$&$a_3$ (\AA K$^{-3})$\\
\hline
classical MD    &  4.3483   & 2.1 10$^{-5}$& 6.4 10$^{-9}$ &-2.3 10$^{-12}$ \\
\hline
\end{tabular}
\label{tab1}
\end{table}
% -----------------------------------------------------------------

It is interesting to quantify the renormalization of the cell parameter
$a(T)$ as a consequence of the quantum character of the atomic nuclei by
comparison to results of classical simulations.
Moreover, the employed formalism allows us to 
treat the atomic nuclei either quantum mechanically or in the classical
limit just by tuning the nuclear mass.
Thus, we have also calculated the effect in $a(T)$ of
considering only one type of nuclei classically.
The results of these simulations are presented in Fig. \ref{fig2}.
The full PI MD results (circles) are compared to classical MD
simulations (triangles).
While the classical calculations yield a finite linear expansion for $T$
tending to zero, the quantum PI MD calculations yield cell parameters
which become independent of $T$ at low $T$, in agreement with the experimental 
results. 
The quantum PI MD results that selectively treat the Si or
the C nuclei as classical particles are shown by squares and diamonds,
respectively. 
The lines represent numerical fits to the simulation results.
The classical MD data were fitted with a cubic function,
\begin{equation}
a_{cla}(T) = \sum_{i=0}^{3} a_i T^i \,\, ,
\label{fit_cla}
\end{equation}
while the PI MD simulation results treating either 
the Si or C nuclei classically
were fitted to Eq. (\ref{fit}) plus an additional linear term,
$a_1T$, to account for the finite slope of the curves at $T=0$.
The values of the fitted coefficients are summarized in Table \ref{tab1}.

The cell parameter $a_{cla}(T=0)$ amounts
to 4.3483 \AA\, while the extrapolated value of the  PI MD
simulation is $a(T=0)$= 4.3594 \AA.
Therefore the zero-point renormalization of the cell parameter of 3C-SiC 
amounts to $\Delta a/a = 2.5\times10^{-3}$ ($\Delta a=a-a_{cla}$).
This value is slightly lower than the average ($2.9\times10^{-3}$)
of the zero-point renormalization 
of diamond ($3.9\times10^{-3}$) 
and Si ($1.9\times10^{-3}$)
derived from experiments on crystals with different 
isotopic masses.\cite{ca05r}

\begin{figure}
\vspace{-3.8cm}
\includegraphics[width= 9cm]{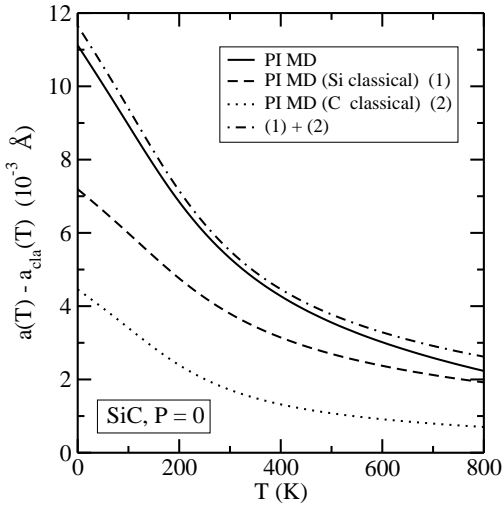}
\vspace{-1.5cm}
\caption{
Temperature dependence of the cell parameter renormalization of
3C-SiC. The results derived from full quantum simulation (continuous line)
are compared to those derived by treating either the Si or the C nuclei
in the classical limit (dashed and dotted lines, respectively).
The dashed-dotted line shows the sum of the separate Si and C
renormalizations. The curves have been derived from the fits shown
in Fig. \ref{fig2}.
}
\label{fig3}
\end{figure}
 
The $a(T)$ curves where
either the C or Si nuclei are treated classically provide
evidence for the linear character of the renormalization of 
the lattice parameter.
This fact is clearly seen in Fig. \ref{fig3},
where the absolute values of the cell parameter renormalization, 
$\Delta a(T)=a(T)-a_{cla}(T)$, are represented 
up to 800 K.
We observe that the sum of the cell parameter renormalizations
obtained when either the Si or C nuclei are treated
quantum mechanically is nearly identical to the total
renormalization obtained in the full PI MD simulations.
This linear behavior is probably related to the fact that the
cell parameter renormalization is relatively small, so that 
in terms of perturbation theory it can be realistically described
by second-order terms in the atomic displacements. 

\begin{figure}
\vspace{-3.8cm}
\includegraphics[width= 9cm]{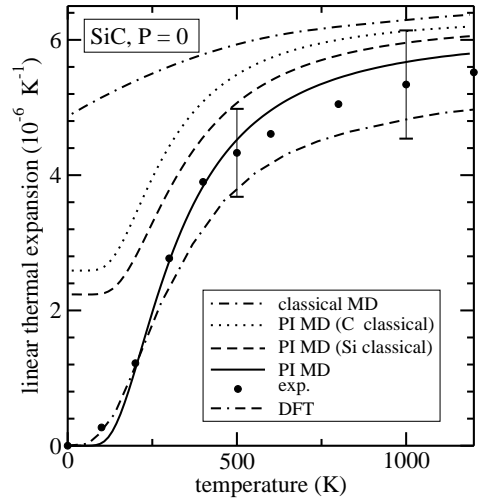}
\vspace{-1.5cm}
\caption{
Thermal expansion coefficient of 3C-SiC as a function of temperature.
Closed circles are experimental data from Ref. \onlinecite{sl75} and
their estimated error bars are shown at 500 and 1000 K.
The simulation results are derived from the numerical fits
shown in Fig. \ref{fig2}.
The classical MD result at temperatures above 800 K requires to
add higher-order terms to the cubic fit given by Eq. (\ref{fit_cla}).
The results derived from DFT calculations in combination with a
quasiharmonic
approximation are taken from Ref. \onlinecite{ka94}.
}
\label{fig4}
\end{figure}

The recommended experimental values of the linear thermal expansion, 
$\alpha(T)$,
of 3C-SiC are shown in Fig. \ref{fig4} as closed circles.
Error bars represent the uncertainty of the experimental results.
The simulation results presented in Fig. \ref{fig4} were derived from the 
fits shown in Fig. \ref{fig2} for $a(T)$.
The agreement between our PI MD results for $\alpha(T)$ and the experimental
data is satisfactory in the whole temperature 
range.
The largest deviation is found at temperatures above 700 K
where the thermal expansion of 3C-SiC is overestimated by about 8\%
by our model.
Although the deviation between PI MD simulated results and
measured data is well within the experimental error bar,
we stress that the TB model seems to overestimate
the anharmonicity of the interatomic potential at temperatures 
above 600 K. The thermal expansion coefficient obtained in the classical
MD simulations at a given temperature is always larger than the
corresponding PI MD value. This behavior is expected from the fact that the 
classical thermal expansion is finite in the zero-temperature limit
(while it vanishes in the quantum case) and from the consideration that 
both sets of simulations should converge one to the other 
at high enough temperatures.
Note that the deviation from experiment is always larger for classical 
than for quantum simulations, while the PI MD results where either the 
Si or the C nuclei are treated as classical particles lie between both 
limits. 
The classical simulation results presented in Fig. \ref{fig4} 
have required to add higher-order terms to the cubic fit
given by Eq. (\ref{fit_cla}) at temperatures above 800 K.

The quality of our results is comparable to previous calculations
of $\alpha(T)$ by using either ab initio DFT electronic Hamiltonians\cite{ka94}
or phenomenological models\cite{tal95} to determine
harmonic vibrational frequencies and a quasiharmonic approximation
to take into account the anharmonicity of the interatomic potential.
The DFT result of Ref. \onlinecite{ka94} has been plotted in Fig. \ref{fig4}.
The deviation from experimental data found at high temperature is probably 
caused by anharmonic effects not taken into account by the quasiharmonic
approximation.

% -----------------------------------------------------------------
\begin{table}
\caption{
Relative one-electron energies at the symmetry point $\Gamma$ for a
static 3C-SiC lattice.
LMTO results are from Ref. \onlinecite{wi95}, and EPM data
from Ref. \onlinecite{he72}.
All values in eV.
}
~\newline
\begin{tabular}{|l|c|c|c|}
\hline
                    &   EPM   &    LMTO   &      TB   \\
\hline
$\Gamma_{15}^c$     &   6.5   &    7.8    &      7.6  \\
$\Gamma_{1}^c$      &   5.9   &    6.7    &      9.9  \\
$\Gamma_{15}^v$     &   0     &    0      &      0    \\
$\Gamma_{1}^v$      & -19.0   &  -15.5    &    -13.6  \\
\hline
\end{tabular}
\label{tab_e}
\end{table}
% -----------------------------------------------------------------

\subsection{Direct electronic gap}

The one-electron energies derived by the TB model
for 3C-SiC at the symmetry point $\Gamma$ are compared to results
of linear muffin-tin-orbital calculations\cite{wi95} 
and empirical pseudopotential method\cite{he72}    
in Table \ref{tab_e}.
These energies are obtained with the atoms fixed in their crystallographic
positions and thus neglect the effects of lattice vibrations.
The employed TB model predicts that the first
direct gap, $E_0$, of 3C-SiC at $\Gamma$ appears 
between electronic states with $\Gamma_{15}$ symmetry.
This fact is in contradiction with the other electronic Hamiltonians
that show that the conduction band bottom 
is of $\Gamma_{1}$ symmetry at $\Gamma$.\cite{wi95,he72}    
In the case of diamond and Si the bottom of the
conduction band at $\Gamma$ is found to be of $\Gamma_{15}$ 
symmetry by ab initio calculations.\cite{pa71,te79}
The repulsion between the $\Gamma_{25'}$ valence band and the $\Gamma_{15}$
conduction band, induced by the asymmetric potential, is probably 
responsible for the lowering of the $\Gamma_1$ conduction band
with respect to the $\Gamma_{15}$ counterpart in SiC.
Although the employed TB model does not provide an accurate description
of the conduction bands, it has demonstrated, by the study of 
the direct gap of diamond, that is a realistic starting model to study  
electron-phonon interaction effects.\cite{ra06}

\begin{figure}
\vspace{-3.8cm}
\includegraphics[width= 9cm]{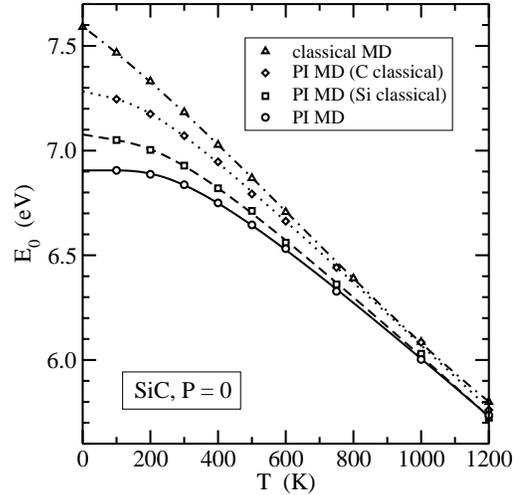}
\vspace{-1.5cm}
\caption{
Temperature dependence of the direct energy gap, $E_0(T)$,
of 3C-SiC at $\Gamma$ at $P=0$.
The results derived from full quantum simulation
(circles) are compared to PI MD simulations
performed by selectively setting the Si nuclei (squares)
or the C nuclei (diamonds) as classical particles.
The results of classical MD simulations are shown by triangles.
The lines represent empirical fits as explained in the text.
The statistical error of the simulation results is less than the
size of the symbols.
}
\label{fig5}
\end{figure}

The temperature dependence of the direct gap, $E_0(T)$, derived by our
$NPT$ simulations of 3C-SiC is shown by circles in Fig. \ref{fig5}.
The continuous line represents a fit of $E_0(T)$ to the
Bose-Einstein expression
\begin{equation}
E_0(T) = e_0 - g \left(1 + \frac{2}{\exp(\Theta_E/T) -1} \right)  \,\,  .
\label{fit_gap}
\end{equation}
The extrapolated value of $E_0(T=0)$ amounts to 6.9 eV, while at
300 K the gap is reduced to 6.84 eV, and at 1000 K amounts to 6 eV. 
The decrease of $E_0$ with temperature is an
effect of the electron-phonon interaction, and therefore
it is expected to depend on the amplitude of the nuclei displacements.
To assess this point, the results obtained for $E_0(T)$ by treating
either the Si or the C nuclei as classical particles
are presented in Fig. \ref{fig5} as squares and diamonds, respectively.
In both cases the results were fitted to
Eq. (\ref{fit_gap}) with an additional linear term, $e_1T$, to account for
the finite slope at $T=0$.
The fitted parameters are summarized in Table \ref{tab2}.
The results of classical MD simulation for 
$E_{0,cla}(T)$ are given by triangles in Fig. \ref{fig5}, while the 
dashed-dotted line is a cubic fit as in 
Eq. (\ref{fit_cla}).
The classical simulation predicts $E_{0,cla}(T=0)=7.59$ eV,
thus the zero-point renormalization of the
direct energy gap, $E_0-E_{0,cla}$, amounts to --0.69 eV, 
which is roughly 10\% of the value of the gap. 
The zero-point renormalization in 3C-SiC
is similar to that found for diamond by either 
PI MD simulations\cite{ra06} or by
perturbation theory.\cite{zo92}

% -----------------------------------------------------------------
\begin{table}
\caption{
Values of the parameters obtained by fitting the temperature
dependence of the direct energy gap, $E_0(T)$,
of 3C-SiC with Eq. (\ref{fit_gap}).
All results were derived by $NPT$ simulations at $P=0$.
The PI MD parameters for those cases where the Si or C nuclei are
treated in the classical limit were obtained by adding a linear term, $e_1T$,
to Eq. (\ref{fit_gap}).
The classical MD parameters correspond to a cubic function as in
Eq. (\ref{fit_cla}).
}
~\newline
\begin{tabular}{|l|c|c|c|c|}
\hline
                &$e_0$ (eV) & $e_1$ (eV K$^{-1})$ & $g$(eV)   &$\Theta_E$(K)\\
\colrule
PI MD           &  7.548    &   -          & 0.643        &     887.2        \\
PI MD (Si cla.) &  7.578    &-2.7 10$^{-4}$& 0.501        &     817.9        \\
PI MD (C  cla.) &  7.646    &-3.6 10$^{-4}$& 0.363        &     615.7        \\
\hline
                &$e_0$ (eV) & $e_1$ (eV K$^{-1}$)& $e_2$ (eV K$^{-2})$&$e_3$
(eV K$^{-3})$\\
\hline
classical MD    &  7.593    &-12.1 10$^{-4}$&-6.2 10$^{-7}$ & 3.2 10$^{-10}$ \\
\hline
\end{tabular}
\label{tab2}
\end{table}
% -----------------------------------------------------------------

\begin{figure}
\vspace{-3.8cm}
\includegraphics[width= 9cm]{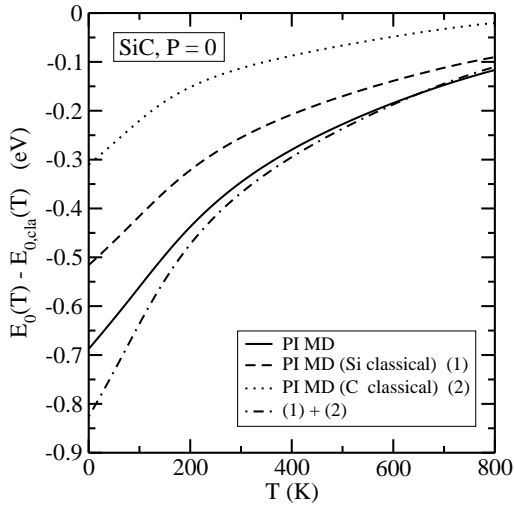}
\vspace{-1.5cm}
\caption{
Temperature dependence of the direct band gap renormalization,
$\Delta E_0(T)$, of 3C-SiC at $P=0$.
The results derived from full quantum simulation (continuous line)
are compared to those derived by treating either the Si or the C nuclei
in the classical limit (dashed and dotted lines, respectively).
The dashed-dotted line shows the sum of the separate Si and C
renormalizations. The curves have been derived from the fits shown
in Fig. \ref{fig5}.
}
\label{fig6}
\end{figure}

The temperature dependence of the direct gap renormalization,
$\Delta E_0(T) =  E_0(T) -  E_{0,cla}(T)$ is shown
in Fig. \ref{fig6} by a full line.
The separate Si contribution to the  gap renormalization is obtained
by a simulation where the C nuclei are treated classically. 
This result is shown by a dotted line in Fig. \ref{fig6}.
The dashed line represents the separate C contribution as derived
from a simulation with classical Si nuclei.
The sum of both Si and C increments (dashed-dotted line) is somewhat larger
than the gap renormalization obtained in the full quantum
simulation of 3C-SiC, in particular at temperatures below 250 K,
where the gap renormalization reaches its largest values.
This non-linear behavior of $\Delta E_0(T)$ is in contrast to the linear one 
found for the lattice parameter renormalization, $\Delta a(T)$,
in Fig. \ref{fig3}.
This non-linearity is probably related to the fact that 
the relative gap renormalizations
are found to be much larger than the relative cell parameter renormalizations.
In terms of perturbation theory, the fact that
the total gap renormalization is lower that the
separate Si and C contributions, implies that fourth-order
terms in the nuclei displacements are important
for $\Delta E_0(T)$, and their contribution is of opposite sign
to the leading second-order terms.

\begin{figure}
\vspace{-3.8cm}
\includegraphics[width= 9cm]{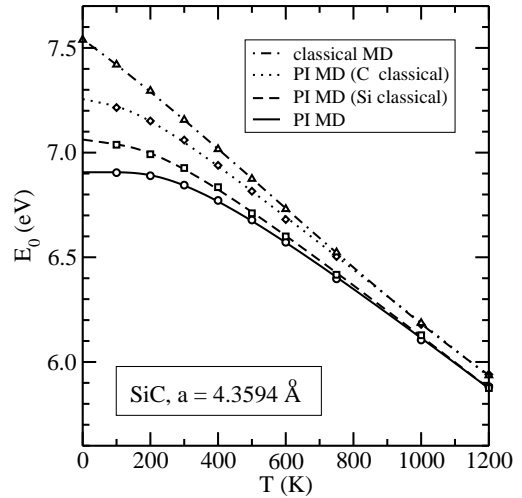}
\vspace{-1.5cm}
\caption{
Temperature dependence of the direct energy gap, $E_0(T)$,
of 3C-SiC at $\Gamma$ at constant volume.
The results derived from full quantum simulation
(circles) are compared to PI MD simulations
performed by selectively setting either the Si nuclei (squares)
or the C nuclei (diamonds) as classical particles.
The results of classical MD simulations are shown by triangles.
The lines represent empirical fits as explained in the text.
The statistical error of the simulation results is less than the
size of the symbols.
}
\label{fig7}
\end{figure}

% -----------------------------------------------------------------
\begin{table}
\caption{
Values of the parameters obtained by fitting the temperature
dependence of the direct energy gap, $E_0(T)$, of 3C-SiC.
The results correspond to constant volume $NVT$ simulations using
a cell parameter of $a=4.3594$ \AA.
See Tab. \ref{tab2} for details on the fitting functions.
}
~\newline
\hspace*{-0.5cm}
\begin{tabular}{|l|c|c|c|c|}
\hline
                &$e_0$ (eV) & $e_1$ (eV K$^{-1})$ & $g$(eV)   &$\Theta_E$(K)\\
\colrule
PI MD           &  7.461    &   -          & 0.555        &     876.4        \\
PI MD (Si cla.) &  7.442    &-2.3 10$^{-4}$& 0.379        &     722.6        \\
PI MD (C  cla.) &  7.508    &-3.2 10$^{-4}$& 0.252        &     517.5        \\
\hline
                &$e_0$ (eV) & $e_1$ (eV K$^{-1}$)& $e_2$ (eV K$^{-2})$&$e_3$
(eV K$^{-3})$\\
\hline
classical MD    &  7.541    &-11.8 10$^{-4}$&-4.2 10$^{-7}$ & 2.4 10$^{-10}$ \\
\hline
\end{tabular}
\label{tab3}
\end{table}
% -----------------------------------------------------------------

\begin{figure}
\vspace{-3.8cm}
\includegraphics[width= 9cm]{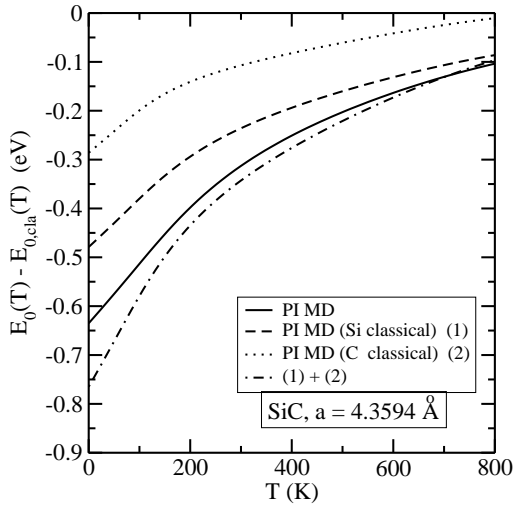}
\vspace{-1.5cm}
\caption{
Temperature dependence of the direct band gap renormalization,
$\Delta E_0(T)$, of 3C-SiC at constant volume.
The results derived from full quantum simulation (continuous line)
are compared to those derived by treating either the Si or the C nuclei
in the classical limit (dashed and dotted lines, respectively).
The dashed-dotted line shows the sum of the separate Si and C
renormalizations. The curves have been derived from the fits shown
in Fig. \ref{fig7}.
}
\vspace{1.5cm}
\label{fig8}
\end{figure}

The conclusion that fourth-order terms in the gap renormalization
are not related to changes in the cell parameter is further
demonstrated by the results shown in Figs. \ref{fig7} and
\ref{fig8}, where we show the values of the direct gap, $E_0(T)$,
and gap renormalizations, $\Delta E_0(T)$, 
derived at constant volume by $NVT$ simulations.
The volume was kept fixed at the equilibrium value of the cell
parameter ($a=4.3594$ \AA) extrapolated for $T=0$ from 
our PI MD simulations (see Fig. \ref{fig1}).
The direct gap, $E_0(T)$, in Fig. \ref{fig7} is
presented for the full PI MD simulation, the classical  
MD simulation, and PI MD simulations treating either the Si or the C nuclei
as classical particles.
For each case numerical fits to the simulation results were
performed in the same way as explained for the 
zero-pressure results in Fig. \ref{fig5}.
The fitted parameters are summarized in Table \ref{tab3}. 
The gap renormalizations, $\Delta E_0(T)$, in Fig. \ref{fig8}
show that non-linear effects appear again at temperatures below 250 K
in the case of constant volume simulations.
Note that the comparison between Figs. \ref{fig6} and \ref{fig8} 
reveals that non-linear effects in the gap renormalization
are independent of the volume.

\begin{figure}
\vspace{-0.0cm}
\includegraphics[width= 9cm]{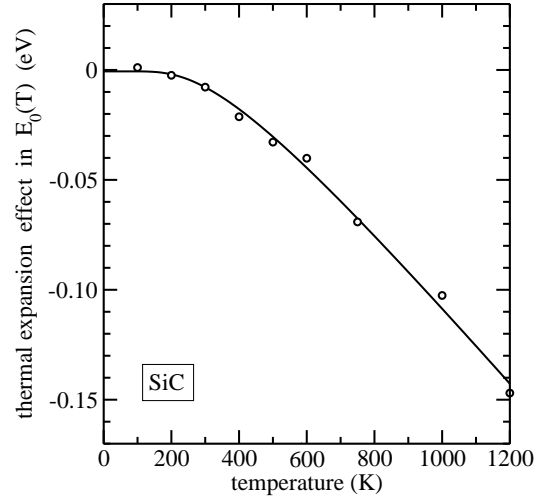}
\vspace{-1.5cm}
\caption{
Thermal expansion effect in the
value of the direct electronic gap, $E_0$,
of 3C-SiC as a function of temperature.
The circles are the difference between the $E_0(T)$ values
obtained by PI MD simulations at $P=0$ (see Fig. \ref{fig5})
and at constant volume (see Fig. \ref{fig7}).
The line is a guide to the eye.
}
\vspace{1.5cm}
\label{fig9}
\end{figure}

The difference between the values of $E_0(T)$ obtained in our PI MD 
simulations at $P=0$ (Fig. \ref{fig5}) and at constant volume (Fig. \ref{fig7})
allows us to quantify the thermal expansion effect in the gap.
This difference is shown in Fig. \ref{fig9} as a function of temperature.
We note that the thermal expansion produces a 
decrease in the value of $E_0(T)$ as temperature increases.
At 300 K this decrease amounts to only 8 meV, while 
at 1000 K, the decrease rises
to a value slightly less than 110 meV.
These values should be compared to the total effect of
temperature in the direct gap,
$E_0(T) - E_0(T=0)$, which amounts to 64 meV at 300 K
and to 900 meV at 1000 K, as derived from the PI MD results
shown in Fig. \ref{fig5}.
Thus, the thermal expansion is responsible for
about 10 \% of the total decrease in the value of $E_0$
at temperatures of 300 and 1000 K.

\subsection{Comparison of $E_0$ for diamond, 3C-SiC and Si}

It is interesting to compare the simulation results of the direct gap
$E_0(T)$ found for 3C-SiC with those corresponding to diamond and Si
by using the same tight-binding parameterization. 
The studied direct gap, $E_0$, for both diamond and Si
corresponds to transitions between one-electron states with
symmetry $\Gamma_{25'}$ (valence band) and $\Gamma_{15}$
(conduction band).
To prevent possible inaccuracies of the TB model in the determination
of the thermal expansion of diamond and Si, we have performed
constant volume simulations of diamond and Si with the following
values of the cell parameters, $a_C=3.567$ \AA\ for diamond\cite{ka59}
and $a_{Si}=5.430$ \AA\ for Si.\cite{re75} 
Our simulation results of $E_0(T)$ in diamond, 3C-SiC, and Si
will be also compared to available experimental data.
Note that in this comparison the thermal expansion effect in 
$E_0(T)$ is not included in our constant volume simulations. 

\begin{figure}
\vspace{-0.0cm}
\includegraphics[width= 9cm]{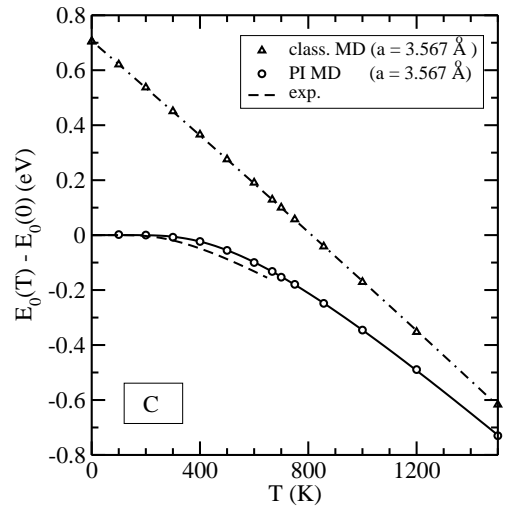}
\vspace{-1.5cm}
\caption{
Relative shifts of the direct
electronic gap of diamond obtained by
our PI MD (circles) and classical MD simulations (triangles)
as a function of temperature.
The dashed line is a fit to the experimental data of
Logothetidis {\it et al.} for diamond IIa.\cite{lo92}
The zero of the energy scale was set at 7.06 eV for both experimental
and simulation results.
}
\label{fig10}
\end{figure}

In Fig. \ref{fig10} we show the results of the PI MD and classical
MD simulations for the relative shifts of the
direct gap of diamond with temperature. 
The lines through the points correspond to numerical
fits using Eqs. (\ref{fit_gap}) and (\ref{fit_cla}), respectively.
The fitted parameters are summarized in Table \ref{tab4}.
The zero-point renormalization of $E_0$ amounts to 0.705 eV,
a value that agrees well with the result of 0.678 eV 
derived by Zollner {\it et al.} in Ref. \onlinecite{zo92}
by a perturbational treatment of the electron-phonon coupling.
Unfortunately there is no experimental estimation of the renormalization
of the direct gap, $E_0$, of diamond.
However, the zero-point renormalization of the {\it indirect} gap of diamond
derived from luminescence data amounts to 0.37 eV.\cite{ca05} 
The broken line shown in Fig. \ref{fig10} is the experimental result reported
in Ref. \onlinecite{lo92} for diamond IIa, based on
measurements of the complex dielectric function by spectroscopic
ellipsometry between 100 to 650 K.
The extrapolated experimental value $E_0(T=0)$ varies
from 7.06 to 7.14 eV, depending on the line-shape analysis
of the spectra by using a first or a second derivative.
The TB model gives a value of $E_0(T=0)=7.06$ eV in good
agreement with experiment.
The slope of the PI MD results at temperatures above 500 K
is larger than that of the experimental data, a fact that
indicates that the employed TB model overestimates the
electron-phonon interaction at temperatures above 500 K.

% -----------------------------------------------------------------
\begin{table}
\caption{
Values of the parameters obtained by fitting the temperature
dependence of the direct energy gap, $E_0(T)$, of diamond,
3C-SiC, and Si with Eq. (\ref{fit_gap}).
The PI MD results correspond to constant volume $NVT$ simulations.
}
~\newline
\begin{tabular}{|l|c|c|c|}
\hline
                             &$e_0$ (eV)& $g$(eV)      &  $\Theta_E$(K)   \\
\colrule
PI MD C  ($a$=3.567  \AA)    &  7.802   & 0.740        &    1665.0      \\
PI MD 3C-SiC ($a$=4.3594 \AA)&  7.461   & 0.555        &     876.4      \\
PI MD Si ($a$=5.4296 \AA)    &  3.511   & 0.204        &     513.6      \\
exp. C\footnote{Logothetidis {\it et al.}, Ref. \onlinecite{lo92}}
                             &  7.387   & 0.320        &    1060.0      \\
exp. 3C-SiC\footnote{Petalas {\it et al.}, Ref. \onlinecite{pe98}}
                             &  7.943   & 0.230        &     668.0      \\
exp. Si\footnote{Jellison and Modine, Ref. \onlinecite{je83}}
                             &  3.467   & 0.091        &     382.6      \\
exp. Si\footnote{Lautenschlager {\it et al.}, Ref. \onlinecite{la87}}
                             &  3.378   & 0.025        &     267.0      \\
\hline
\end{tabular}
\label{tab4}
\end{table}
% -----------------------------------------------------------------

\begin{figure}
\vspace{-0.5cm}
\includegraphics[width= 9cm]{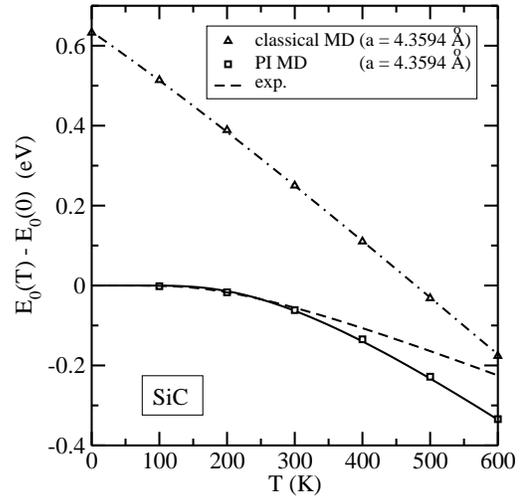}
\vspace{-1.5cm}
\caption{
Relative shifts of the direct
electronic gap of 3C-SiC obtained by
our PI MD (circles) and classical MD simulations (triangles)
as a function of temperature.
The dashed line is a fit to the experimental data of
Petalas {\it et al.}\cite{pe98}
The zero of the energy scale corresponds to 6.91 eV for the
simulation results and to 7.61 eV for the experimental ones.
}
\vspace{1.5cm}
\label{fig11}
\end{figure}

In Fig. \ref{fig11} we compare the relative shifts of the direct energy
gap of 3C-SiC obtained by our PI MD and classical MD simulations
(circles and triangles, respectively) to a fit to the experimental
data derived between 90 and 550 K by spectroscopic ellipsometry.
The experimental data can not discriminate
between interband electronic transitions occurring at the points
$\Gamma$ (critical point $E_0$), and along $\Lambda$ (critical point $E_1$) 
in reciprocal space, 
as they appear in the same energy region. 
The fitted parameters are summarized in Table \ref{tab4}.
The comparison to experiment shows that
our computational model tends to overestimate the 
shift in the energy gap.

\begin{figure}
\vspace{-0.0cm}
\includegraphics[width= 9cm]{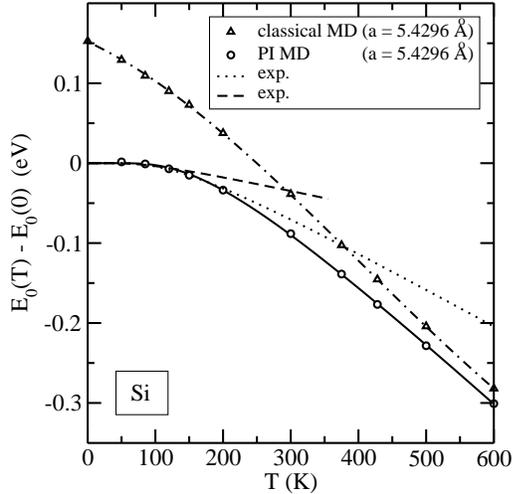}
\vspace{-1.5cm}
\caption{
Relative shifts of the direct
electronic gap of Si obtained by
our PI MD (circles) and classical MD simulations (triangles)
as a function of temperature.
The dashed line is a fit to the experimental data of
Lautenschlager {\it et al.},\cite{la87}
while the dotted line is a fit to the Jellison and Modine
results based on polarization
modulation ellipsometry.\cite{je83}
The zero of the energy scale is set at 3.31 eV for the simulation
results, at 3.35 eV for the dashed line and at 3.38 eV for the
dotted line.
}
\vspace{1.5cm}
\label{fig12}
\end{figure}

The simulation results for the temperature shift of the
direct electronic gap of Si are compared to available 
experimental data in Fig. \ref{fig12}.
The dashed line represents the numerical fit to the
spectroscopic ellipsometric data of
Lautenschlager {\it et al.},\cite{la87}
while the dotted line is derived by polarization modulation
ellipsometry.\cite{je83}
The difference between both sets of experimental data might
be due to the fact that the measured excitations are a 
superposition of interband transitions ($E_0$, $E_1$) along the
$\Lambda$ direction that includes both the $\Gamma$
and $L$ points at its boundary.
Fitted parameters are collected in Table \ref{tab4}.
The extrapolated PI MD value of $E_0(T=0)$ amounts to 3.31 eV,
in reasonable agreement to the experimental extrapolated 
results of 3.35 eV,\cite{la87} and 3.38 eV.\cite{je83} 
The calculated zero-point renormalization of $E_0$ amounts to 
0.15 eV.
This result is to be compared to the experimental value of
0.12$\pm$0.02 eV derived by Lastras-Mart\'{\i}nez {\it et al.} 
from a study of isotopically pure and natural Si.\cite{la00}
This experimental value has to be considered as a 
mixture of $E_0$ and  $E_1$ transitions. 
The temperature shifts in $E_0$ derived from our PI MD 
simulations appear again to be larger than in the experimental
data, pointing toward an overestimation of the electron-phonon interaction
by the employed electronic TB Hamiltonian.

\begin{figure}
\vspace{-0.0cm}
\includegraphics[width= 9cm]{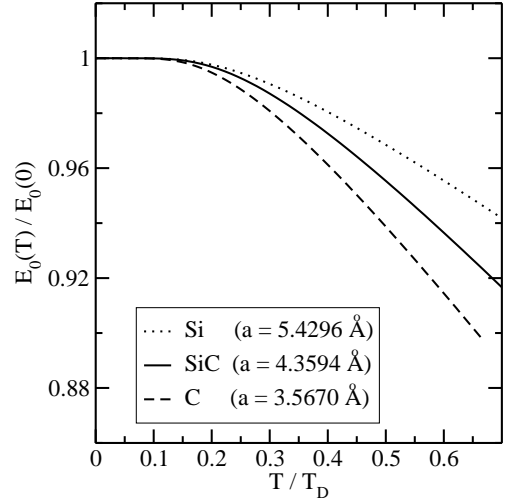}
\vspace{-1.6cm}
\caption{
Relative values of the direct electronic gap of
diamond, 3C-SiC, and Si as a function of reduced temperature.
The Debye temperature, $T_D$,  is 2240 K (diamond),
1080 K (3C-SiC), and 645 K (Si).\cite{sl75}
The results correspond to PI MD simulations at constant
volume.
}
\vspace{2.0cm}
\label{fig13}
\end{figure}

In order to compare the effects of the electron-phonon interaction
in the direct gap, $E_0$, of diamond, 3C-SiC, and Si, we have 
plotted in Fig. \ref{fig13} the simulation results of $E_0(T)$ 
using relative energy and temperature scales.
For each crystal the energy was measured in units of its
direct gap $E_0(T=0)$ and the temperature in units of the corresponding
Debye temperature.
We see that by using reduced units the direct gap of 3C-SiC 
falls roughly between the values found for diamond and Si
in the studied temperature range.
It has been recently shown that at very low temperatures (below 10 K
for Si) the gap changes with $T$ like $T^4$.\cite{ca04}
Our calculations, however, do not have the necessary accuracy
to reveal this dependence.

\subsection{Pressure dependence of $E_0$ in 3C-SiC}

\begin{figure}
\vspace{-0.0cm}
\includegraphics[width= 9cm]{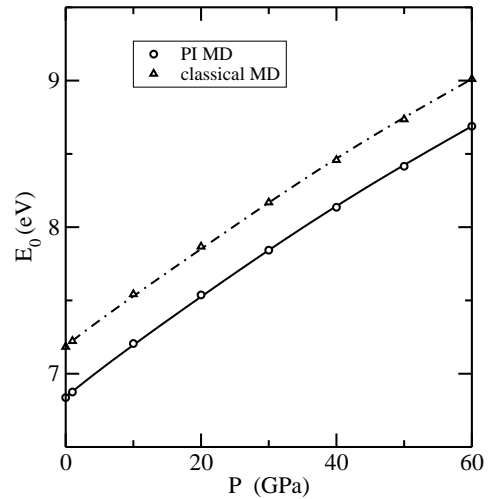}
\vspace{-1.6cm}
\caption{
Pressure dependence of the direct electronic gap of 3C-SiC
as derived from PI MD and classical MD simulations
in the $NPT$ ensemble at 300 K.
The statistical error of the simulation results is less than the
size of the symbols.
}
\vspace{1.3cm}
\label{fig14}
\end{figure}

The pressure dependence of the direct gap of
3C-SiC at 300 K was derived from simulations in the
$NPT$ ensemble up to 60 GPa.
The results of the quantum PI MD and classical simulations
are shown in Fig. \ref{fig14}.
The main difference between both sets of
results is a rigid shift of the $E_0$ values,
that reflects the dependence of the electron-phonon
coupling on the nuclei displacements around the equilibrium positions.
At a given pressure, the vibrational amplitudes 
are always larger for the quantum simulations.
The cell parameter difference, $a$ versus $a_{cla}$,
has a relatively smaller effect in the shift of $E_0$.
For example, at $P=0$ the quantum result for the direct
gap is 0.35 eV lower
than the classical one (see Fig. \ref{fig14}), a value that includes
the effect of the volume difference in the quantum and classical 
simulations.
The corresponding result obtained at a constant volume at 300 K
is 0.31 eV (see Fig. \ref{fig7}).

The derivative, $dE_0/dP$, at $P=0$ is readily obtained from the data
in Fig. \ref{fig14}, giving a result of 39 meV/GPa in the quantum case
versus a value of 41 meV/GPa in the classical limit at 300 K.
These values are in reasonable agreement with the calculation
of Park {\it et al.},\cite{pa94} which gives a value of 51 meV/GPa,
based on ab initio calculations using a local-density-functional
approximation (LDA) without considering any kind of temperature effect.

\begin{figure}
\vspace{-0.0cm}
\includegraphics[width= 9cm]{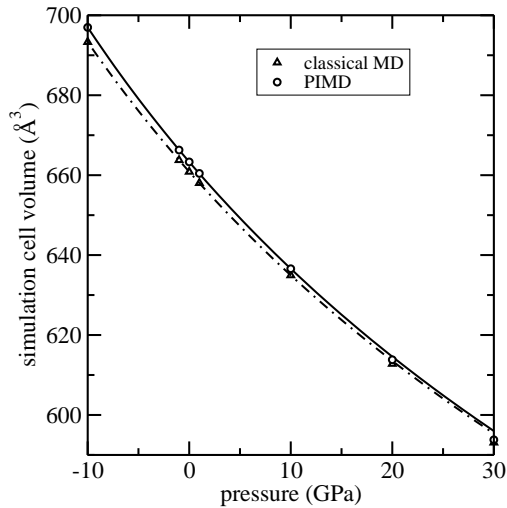}
\vspace{-1.5cm}
\caption{
$P-V$ curves derived for 3C-SiC
PI MD and classical MD simulations
in the $NPT$ ensemble at 300 K.
The lines are fits to the Murnaghan's equation of state
using the points in the interval between [-10,10] GPa.
The statistical error of the simulation results is of the
size of the symbols.
}
\label{fig15}
\end{figure}

Finally, the set of simulations performed at 300 K 
allows us to plot $P-V$ curves 
that can be used to derive the value of the bulk modulus, $B_0$,
and its pressure derivative, $B_0'$, for 3C-SiC.
In Fig. \ref{fig15} we show the $P-V$ curves obtained from
$NPT$ simulations of 3C-SiC for several pressures in the range
between -10 and 30 GPa.
The five points calculated in the pressure interval [-10,10] GPa
were used to fit a Murnaghan equation of state\cite{mu44}
\begin{equation}
V = V_0 \left( B_0'\frac{P}{B_0} + 1\right)^{-1/B_0'}   \,\,  ,
\end{equation}
where the subindex 0 indicates values at $P=0$.
The continuous line in Fig. \ref{fig15} shows the fit to the
PI MC simulations, that provides the values
$B_0$=225 GPa and $B_0'=4.1$.
We checked that this value of $B_0$ is consistent with that
($B_0=221\pm8$ GPa) derived by calculating the volume fluctuations in 
the $NPT$ ensemble at $P=0$. The fluctuation relation for a given pressure
is    
\begin{equation}
B = k_B T \frac{\langle V \rangle}
                 {\langle V^2 \rangle - \langle V \rangle^2}    \,\,  .
\end{equation}      
The calculated value of $B_0$ is in excellent
agreement with the experimental data,\cite{ag06,la91,ca68}
and also the calculated value of $B_0'=4.1$ agrees with 
experiment ($B_0'=4\pm0.3$).\cite{st87}
The values obtained from the Murnaghan fit of the classical
simulation results were $B_{0,cla}$=230 GPa and $B_{0,cla}'=4.2$.

\section{Conclusions}

The simulation method employed in this work has demonstrated its
capability for the description of anharmonic effects related to 
the phonon-phonon interaction, as well as for the treatment of 
the electron-phonon coupling in semiconducting
solids as a function of temperature.
Thus, this type of simulations is an alternative to perturbational
treatments, with the advantage of being also applicable in cases
where the convergence of a perturbational series might be slow.
A prerequisite to account for phonon-phonon and electron-phonon
interactions is a quantum description that includes both electrons
and nuclei. 
In this respect, the Feynman formulation of statistical mechanics
allows us to simulate the quantum mechanical properties of the
atomic nuclei in the solid at finite temperatures with either
Monte Carlo or Molecular Dynamics techniques.
A particular advantage of PI MD is that the algorithm can 
be easily parallelized, as the calculation of total energies and
forces for each of the $L$ different replicas of the solid, 
that are generated by
the discretization of the path integral (Trotter number),
can be run in a different processor.
The use of an electronic Hamiltonian to describe the interatomic
interactions in the solid allows us to study also electronic 
properties in the simulations.
At present we have limited ourselves to simplified, although 
accurate, Hamiltonians of the tight-binding type.
But in the future, an interesting improvement will be the  
combination of PI with ab initio parameter-free Hamiltonians.

The PI MD simulation of 3C-SiC has been able to reproduce
different experimental properties of the solid.
Moreover, the comparison to classical simulations 
allowed us to evaluate the magnitude of quantum effects
as a function of temperature, as well as to determine the
separate contribution of each type of nuclei (either C or Si)
to several properties.
The employed  potential model predicts a cell parameter as a 
function of temperature, $a(T)$, in good agreement with the
available experimental data.
The correct description of this anharmonic property points toward
a realistic treatment of the phonon-phonon interactions by
our computational model.
The main discrepancy is found at
temperatures above 600 K, where the linear expansion coefficient is
predicted to be about 8\% larger than the experimental one. 
The zero-point renormalization of the cell parameter of
3C-SiC is calculated to be $\Delta a/a=2.5\times10^{-3}$ 
($\Delta a=a-a_{cla}$).
We find that this value is close to the sum of zero-point renormalizations
obtained when only one type of atomic nuclei (either Si or C) is treated
quantum mechanically.
This near linear behavior is probably related to the fact that
the cell parameter renormalization depends only on second-order 
terms in the nuclei displacements, due to the small magnitude
of $\Delta a/a$.  
            
The direct electronic gap at $\Gamma$ of 3C-SiC has a significant
temperature dependence as a consequence of the electron-phonon
coupling.
The effect of the zero-point vibrations of the lattice phonons 
leads to a gap renormalization of $\Delta E_0=-0.69$ eV
($\Delta E_0=E_0-E_{0,cla}$).
This effect is so large that any theoretical approach
aiming at a quantitative determination of the direct
electronic gap can not be based only on an improved solution
of the many-body electron problem, but it should also include the treatment
of the electron-phonon interaction.
The calculated relative value of the zero-point renormalization 
is $\Delta E_0/E_0=0.10$. 
In this case, the sum of the separate Si and C contributions 
is found to be lower than the total zero-point renormalization.
This non-linear behavior suggests that
fourth-order terms in the nuclei displacements are important
in this case,
as a result of the large value of $\Delta E_0/E_0$.
The ratio of the partial contributions of C and Si to the zero-point gap
renormalization is 1.67 at $P=0$. This is rather close to the inverse 
square root of the ratio of the corresponding masses 
$(28/12)^{1/2}=1.53$.
The latter represents the ratio of squares of zero-point 
vibrational amplitudes,
under the assumption of equal force constants, in the
harmonic approximation.
The calculated pressure coefficient of $E_0$ is 39 meV/GPa 
at 300 K. An ab initio calculation without considering
temperature and electron-phonon interactions gives a value
of 51 meV/GPa. 
Our calculation of the bulk modulus of 3C-SiC at 300 K ($B=225$ GPa) 
and its pressure derivative ($B'=4.1$) shows quantitative agreement
with experiment.

The calculated zero-point renormalization of the direct gap at $\Gamma$
of diamond and Si amounts to 0.7 eV and 0.15 eV, respectively.
For Si the renormalization derived from spectroscopic ellipsometry of
isotopic crystals amounts to 0.12$\pm$0.02 eV.\cite{la00}
The experiment can not discriminate between 
direct transitions at $\Gamma$ and along $\Lambda$, as they appear at
similar energies.
In the case of diamond, a calculation based on perturbation theory
results in a value of the direct gap renormalization of 0.68 eV,\cite{zo92}
in good agreement with our non-perturbational result.
The experimental value for the renormalization of the 
{\it indirect} gap of diamond amounts to 0.37 eV.\cite{ca05}
The comparison of the simulation results 
of diamond, 3C-SiC and Si with available experimental data 
shows that the employed model Hamiltonian tends to overestimate
the decrease of $E_0$ with temperature, i.e., the electron-phonon
coupling in the employed TB model seems to be too strong.
This fact is another motivation for improving the electronic
model in future work.

\begin{acknowledgments}
The calculations presented here were performed at the Barcelona
Supercomputing Center (BSC-CNS).  
This work was supported by CICYT through Grant
No. FIS2006-12117-C04-03 and by CAM through project S-0505/ESP/000237.
ERH thanks DURSI (regional government of Catalonia)
for funding through project 2005SGR683.
\end{acknowledgments}

%%%%%%%%%%%%%%%%%%%%%%%%%%%%%%%%%%%%%%%%%%%%%%%%%%%%%%%%%%%%%%%%%%%%%%%%%%%
%%%%%%%                       APPENDICES
%%%%%%%%%%%%%%%%%%%%%%%%%%%%%%%%%%%%%%%%%%%%%%%%%%%%%%%%%%%%%%%%%%%%%%%%%%%
 
\appendix
\section{Pair potential for Si-C}
\label{app1}   
The pair-potential is parameterized by using the following
functional form\cite{po95}

\begin{equation} 
E_{SiC}(R) = \sum_{n=2}^{10} d_n (R_c-R)^n  
\end{equation} 
where $R$ is the interatomic Si-C distance, $R_c$=4.4 au, 
and $E_{SiC}(R)=0$ if $R > R_c$. The original coefficients $d_n$ were 
modified to decrease the anharmonicity of the Si-C potential
at distances around 3.55 au, that corresponds to the nearest-neighbors
in 3C-SiC. The modified $d_n$ coefficients are (in au): 
$d_2   = 0.06825$,
$d_3   =-0.49329$,
$d_4   = 2.37716$,
$d_5   =-5.79511$,
$d_6   = 7.90779$,
$d_7   =-6.27467$,
$d_8   = 2.86625$,
$d_9   =-0.69579$,
$d_{10}= 0.06926$.
The procedure to fix the $d_n$ coefficients was to calculate curves of 
internal energy versus volume in the classical limit at $T=0$,
i.e., the atoms occupy fixed equilibrium positions at a given volume. 
The Taylor expansion of the internal energy around its minimum
at $V_{cla}$
can be expressed as
\begin{equation} 
\Delta E \approx   
   \frac{1}{2}\frac{B_{cla}}{V_{cla}}          ({\Delta V})^2
 + \frac{1}{6}\frac{B_{cla}}{V_{cla}^2}(-1-B_{cla}') ({\Delta V})^3  \,\, , 
\label{Taylor}
\end{equation}  
with $\Delta V = V - V_{cla}$, and $B_{cla}$, and $B_{cla}'$ being the 
bulk modulus and its pressure derivative in the classical limit at
$T=0$ and $P=0$.
The values used in this expansion were
$B_{cla}$ = 245 GPa, $B_{cla}'$ = 4, 
and $a_{cla}=V_{cla}^{1/3}=4.3457$ \AA. 
The original analytic form of $E_{SiC}(R)$ was modified to 
obtain a reasonable approximation to the $\Delta E$ curve 
in Eq. (\ref{Taylor}) for a range of volumes defined by a cell parameter
in the interval $a_{cla}\pm0.4$ au.

\section{Pressure estimator in the $NVT$ ensemble}
\label{app2}   

The cartesian coordinates of the $N$ atomic nuclei in the crystal
are denoted as $x_i^{(\alpha)}$, where the superscript $\alpha$ runs from
1 to $3N$. The subindex $i$ denotes the ``bead'' associated to a given 
atomic nucleus and runs from 1 to the Trotter number $L$.
The staging coordinates $u_i^{(\alpha)}$ are defined by a linear
transformation of $x_i^{(\alpha)}$ that diagonalizes the harmonic energy
between neighboring beads\cite{tu98,tu02}
\begin{equation} 
u_1^{(\alpha)} = x_1^{(\alpha)}  \,\, ,  
\end{equation}
\begin{equation} 
u_i^{(\alpha)} = x_i^{(\alpha)}   
               - \frac{(i-1)}{i}x_{i+1}^{(\alpha)}   
               - \frac{1}{i}x_{1}^{(\alpha)}   
                        \,\, ,  \,\, i=2,\dots,L \,\, .
\end{equation}
The pressure estimator in our $NVT$ PI MD simulation is  obtained as
\begin{equation}  
{\cal P} = \frac{1}{3V} \sum_{i=1}^{L} 
                       \sum_{\alpha=1}^{3N} 
                             {m_i^{(\alpha)}}
                             {\left[v_i^{(\alpha)}\right]^2}
         - \frac{2}{3V}E_{arm}
         - \frac{1}{L} \sum_{i=1}^{L} 
                         \frac{\partial U({\bf R}_i)}
                              {\partial V}      \,\, ,
\label{P}
\end{equation}
where $m_i^{(\alpha)}$ and $v_i^{(\alpha)}$ represent
the dynamic mass and velocity
associated to the staging coordinate $u_i^{(\alpha)}$. 
These masses are given as\cite{tu98,tu02} 
\begin{equation} 
m_1^{(\alpha)} = m_{\alpha} \,\, ,
\end{equation}
\begin{equation} 
m_i^{(\alpha)} = \frac{i}{i-1} m_{\alpha} \,\, ,  \,\, i=2,\dots,L \,\, ,
\end{equation}
where $m_{\alpha}$ is the nuclear mass associated to coordinate $\alpha$.
$E_{arm}$ represents
the harmonic energy between beads, that in terms of the stagging coordinates
is written as
\begin{equation} 
E_{arm} =  \frac{1}{2} \sum_{i=2}^{L}  
                       \sum_{\alpha=1}^{3N}   
                                     m_i^{(\alpha)} 
                                      {\left[u_i^{(\alpha)}\right]^2} \,\, , 
\end{equation} 
The last summand in Eq. (\ref{P}) represents the volume 
derivative of the potential
energy of the crystal for the nuclei configuration 
${\bf R}_i = (x_i^{1}, \dots, x_{i}^{3N})$. 
We have performed this derivative numerically by considering 
an expansion of $10^{-4}$ \AA\ in the cell parameter, $2a$,
of the simulation cell, although it is also possible to calculate this
derivative from the stress tensor, obtained through the Hellman-Feynman
theorem. The fractional coordinates, 
$x_i^{(\alpha)}/2a$, of the nuclei remain constant along this volume 
expansion, implying that all cartesian coordinates, $x_i^{(\alpha)}$,
must change according to the modified value of $2a$.

%%%%%%%%%%%%%%%%%%%%%%%%%%%%%%%%%%%%%%%%%%%%%%%%%%%%%%%%%%%%%%%%%%%%%%%%%%%
%%%%%%%                       BIBLIOGRAPHY
%%%%%%%%%%%%%%%%%%%%%%%%%%%%%%%%%%%%%%%%%%%%%%%%%%%%%%%%%%%%%%%%%%%%%%%%%%%

\end{document}